\documentclass[reprint,amsmath,amssymb,aps,showpacs]{revtex4}
\usepackage{graphicx}
\usepackage{dcolumn}
\usepackage{bm}
\usepackage{float}
\usepackage{epstopdf}
\usepackage{amsmath}
\usepackage{amsfonts}
\usepackage{amssymb}
\date{today}
\newcommand{\be}{\begin{equation}}
\newcommand{\ee}{\end{equation}}
\newcommand{\bea}{\begin{eqnarray}}
\newcommand{\eea}{\end{eqnarray}}

\newcommand{\la}{\langle}
\newcommand{\ra}{\rangle}

\newcommand{\bS}{{\bf S}}

\newcommand{\bcr}{\begin{array}{clcr}}
\newcommand{\ecr}{\end{array}}

\input{epsf}

\begin{document}

\title{Magnetic phase diagram of a spin  $S=1/2$ antiferromagnetic
two-leg ladder in the presence of modulated along legs
Dzyaloshinskii-Moriya interaction}
\author{N. Avalishvili$^{1,3}$}
\author{B. Beradze$^{2}$}
\author{G.I. Japaridze$^{1,3}$}

\affiliation{\small{\sl $^1$Ilia State University, Faculty of
Natural Sciences and Medicine, 0162, Tbilisi, Georgia}}
\affiliation{\small {\sl $^2$ Ivane Javakhishvili Tbilisi State University,
Faculty of Exact and Natural Sciences, \\
Chavchavadze Avenue 3, 0112, Tbilisi, Georgia}} \affiliation{\small
{\sl $^3$Andronikashvili Institute of Physics, 0177, Tbilisi,
Georgia}}
\date{\today}

\begin{abstract}

We study the ground-state magnetic phase diagram of a spin $S=1/2$
antiferromagnetic two-leg ladder in the presence of period two
lattice units modulated, Dzyaloshinskii-Moriya (DM) interaction
along the legs. We consider the case of collinear DM vectors and
strong rung exchange and magnetic field. In this limit we map the
initial ladder model onto the effective spin $\sigma=1/2$ $XXZ$
chain and study the latter using the continuum-limit bosonization
approach. We identified four quantum phase transitions and
corresponding critical magnetic fields, which mark transitions from
the spin gapped regimes into the gapless quantum spin-liquid
regimes. In the gapped phases the magnetization curve of the system
shows plateaus at magnetisation $M=0$ and to its saturation value
per rung $M=M_{sat}=1$. We have shown that the very presence of
alternating DM interaction leads to opening of a gap in the
excitation spectrum at magnetization $M=0.5M_{sat}$. The width of
the magnetization plateau at $M=0.5M_{sat}$, is determined by the
associated with the dynamical generation of a gap in the spectrum is
calculated and is shown that its length scales as
$(D_{0}D_{1}/J^{2})^{\alpha}$ where $D_{0},D_{1}$ are uniform and
staggered components of the DM term, $J$ is the intraleg exchange
and $\alpha \leq 3/4$ and weakly depends on the DM couplings.

\end{abstract}


\maketitle


\section{The model}\label{sec1}

Low-dimensional quantum magnetism has been the subject of intense
research activity since the pioneering paper by Bethe
\cite{Bethe_31}.  Interest in study of these systems is determined
by their remarkably rich and unconventional low-energy properties
(see for review Refs.
\cite{Mikeska_Kolezhuk_04,Vasiliev_Volkova_18}. An increased current
activity in this field is connected with the large number of
qualitatively new and dominated by the quantum effects phenomena
discovered in these systems
\cite{Sachdev_NP_08,Broholm_et_al_08,Mila_11,Balents_Savary_17} as
well as with the opened wide perspectives for use low-dimensional
magnetic materials in modern nanoscale technologies.

A significant fraction of current research in the field of
low-dimensional magnetism is focused on studies of helical
structures and chiral order in the frustrated quantum magnetic
systems
\cite{Perk_76,Zviagin,Oshikawa_Affleck,YuLu_2003,Aristov_Maleev_00,Tsvelick_01,Zvyagin_Kolezhuk_04,Zvyagin_Kolezhuk_05,Starykh_08,Garate_Affeck_10,Mila_et-al_11,Fazio_14,Starykh_17}.
The key couplings, responsible for stabilization of non-collinear
magnetic configurations in these systems, is the
Dzyaloshinskii-Moriya (DM) interaction \cite{DMI}
\begin{equation}\label{DM-term}
{\cal H}_{DM}=\sum_{n}{\bf D}(n)\cdot[{\bf S}_{n}\times{\bf
S}_{n+1}]\, ,
\end{equation}
where ${\bf D}(n)$ is an axial DM vector. Although generally the
vectors ${\bf D}(n)$ may spatially vary both in direction and
magnitude,  the symmetry  restrictions based on the properties of
real solid state materials usually rule out most  of the
possibilities and  confine  the majority of theoretical discussion
to two principal cases -- uniform DM interaction, ${\bf D}$ vector
remains unchanged over the system
\cite{Zviagin,Aristov_Maleev_00,Tsvelick_01,Starykh_08} and the case
of staggered DM interaction,  with antiparallel orientation of ${\bf
D}$ on adjacent bonds \cite{Oshikawa_Affleck,YuLu_2003}. In both
these cases the DM term can be eliminated by the gauge
transformation and absorbed in a boundary conditions not affecting
the structure of the basic Hamiltonian \cite{Perk_76,Tsvelick_01}.

The spin S=1/2 two-leg ladders represent the another subclass of
low-dimensional quantum magnets which has been also the subject of
perpetual intensive studies during last three decades
\cite{Schulz_86,Affleck_91,Nersesyan_96,Poilblanc_96,Dagotto_Rice_96,Dagotto_99,Mila_98,Totsuka_98,Chitra_Giamarchi_97,Cabra_Honecker_97-98,Usami,GiamarchiTsvelik_99,Wang_2a,Wang_2b,VJM_03,Hida_04,VJM_04,Mila_06,JP_06,JLM_07,JaP_Mah_09,Zheludev_14,Zheludev_15a,Zheludev_15b,Zheludev_16}
. Because the gapped character of the excitation spectrum  of a
two-leg antiferromagnetic ladder for arbitrary ratio of the intraleg
and interleg couplings \cite{Mikeska_Kolezhuk_04}, in the presence
of a magnetic field these systems show dominated by quantum effects
complex behavior, which is manifestly displayed in the magnetic
field driven quantum phase transitions in ladder systems
\cite{Mila_98,Totsuka_98,Chitra_Giamarchi_97,Cabra_Honecker_97-98,Usami,GiamarchiTsvelik_99,Wang_2a,Wang_2b,VJM_04,Mila_06,JP_06,JLM_07}
. During the last few years interest in $S=1/2$ ladder materials has
been enhanced by the discovery of various systems characterized by
the presence of the uniform Dzyaloshinskii-Moriya interaction along
the ladder legs and characterized by unconventional low-temperature
magnetic properties
\cite{Zheludev_14,Zheludev_15a,Zheludev_15b,Zheludev_16}. Among them
are as materials characterized by the strong-leg exchange such as
$(C_7H_10N)_2CuBr_4$ \cite{Zheludev_15a,Zheludev_15b} as well as the
strong rung compounds such as $Cu(C_8H_6N_2)Cl_2$
\cite{Zheludev_16}.

Recently it has been demonstrated that the DM interaction can be
efficiently tailored with an substantial efficiency factor by
structural modulations \cite{Geometric_tailor_DM} or by external
electric field \cite{EF_Enhanc_DM_1,EF_Enhanc_DM_2,EF_Enhanc_DM_3}.
This unveils the possibility to control DM interaction and magnetic
anisotropy via the electric field and opens a wide area for
theoretical consideration of the effects  caused by the spatially
modulated DM interaction on the properties of low-dimensional spin
systems. In the recent publication  it has been shown that the
effect of modulated spatially DMI on the ground state properties of
the spin $S=1/2$ Heisenberg chain is substantial, leading to the
appearance of the additional quantum phase transition point in the
ground state phase diagram, which marks opening of a gap in the
Luttinger-liquid phase of the spin chain and to the formation of the
new gapped phase in the ground state phase diagram characterized by
the coexisting spin dimerization and alternating spin chirality
pattern \cite{AJR_19}.

In the present work we study the effect of the alternating
Dzyaloshinskii-Moriya interaction on the ground state phase
diagram of the spin-1/2 strong-rung Heisenberg ladder in the
presence of magnetic field. The Hamiltonian of the model under
consideration is given by
\begin{equation}\label{Hamiltonian}
\hat{H} = H_{leg}^{1} + H_{leg}^{2} +  H_{rung}\, ,
\end{equation}
where the Hamiltonian for leg $\alpha$ is
\begin{eqnarray}
&&\hspace{-5mm}H_{leg}^{\alpha} = \sum_{n=1}^N
\big\{J{\bS}_{n,\alpha}{\bS}_{n+1,\alpha} + {\bf
D}_{\alpha}(n)\cdot[{\bf S}_{n,\alpha}\times{\bf
S}_{n+1,\alpha}]\big\} \label{Hamiltonian_legs}
\end{eqnarray}
and the term corresponding to the interleg interaction and coupling
with magnetic field is given by
\begin{eqnarray}
H_{rung} &=&  \sum_{n=1}^N \big\{J_{\bot}\,{\bS}_{n,1} \cdot
{\bS}_{n,2} - H\,(S^{z}_{n,1}+S^{z}_{n,2})\big\} .
\label{InterLegCoup}
\end{eqnarray}
Here ${\bS}_{n,\alpha}$ is the spin $S=1/2$ operators at the $n$-th
rung, ${\bf D}_{\alpha}(n)$ is a DM vector and the index
$\alpha=1,2$ denotes the ladder legs. Both the intraleg and interleg
exchange constant are antiferromagnetic ($J,J_{\bot}>0$). In what
follows we restric our consideration by the case where the vectors ${\bf D}_{1}(n)$ and ${\bf
D}_{2}(n)$ are {\em collinear} and the applied magnetic field is parallel to ${\bf D}_{\alpha}(n)$. We take ${\bf D}_{\alpha}(n)=\left(0,0,D_{\alpha}(n)\right)$,
where
\begin{eqnarray}
D_{\alpha}(n)&=&D_{0}^{\alpha}+(-1)^{n}D_{1}^{\alpha}\, .
\end{eqnarray}
It is obvious, that at $D_{\alpha}(n) \neq 0$  the  spin-rotation symmetry of the Hamiltonian
is $U(1)$ and reflects invariance of the system with respect to the rotation around the vector ${\bf D}$ ($\hat{z}$ axis). In addition, at $D_{1}^{\alpha}$  the translational symmetry of the Hamiltonian is restored only after the shift on two lattice units.

The outline of the paper is as follows: In the forthcoming section
we derive transformation which allows to gauge away the DM coupling.
Starting from the section III we restrict our consideration by the
limit of strong rung exchange and magnetic field ($J_{\bot},H \gg
\tilde{J}_{\alpha}$) and derive the effective spin-chain
Hamiltonian.  In the Section IV we use the continuum-limit
bosonization approach to study the ground state properties of the
effective model. Finally, we conclude and summarize our results in
section V.

\newpage
\section{Gauging away the DM interaction} \label{Section2}\,

Because the alternating DM term breaks the translation symmetry, it
is convenient to rewrite the leg Hamiltonian in a form, which
explicitly incorporates doubling of the unit cell of the model. We
define the dimensionless parameters
$d_{\pm}^{\alpha}=(d^{\alpha}_{0} \pm d_{1}^{\alpha})/J$, where
$d^{\alpha}_{0}=D^{\alpha}_{0}/J$ and
$d^{\alpha}_{1}=D^{\alpha}_{1}/J$ and rewrite the $\alpha$-leg
Hamiltonian in the following way
\begin{eqnarray}
\label{Hamilton_XXZ_w_Alt_DMI_2}H_{leg}^{\alpha} &=&\frac{J}{2}
\sum_{m=1}^{N/2}\Big[\,\left(S^{+}_{2m-1,\alpha}S^{-}_{2m,\alpha} +
S^{-}_{2m-1,\alpha}S^{+}_{2m,\alpha}\right)+\left(S^{+}_{2m,\alpha}S^{-}_{2m+1,\alpha}
+
S^{-}_{2m,\alpha}S^{+}_{2m+1,\alpha}\right)\nonumber\\
&+&i\,d^{\alpha}_{-}\left(S^{+}_{2m-1,\alpha}S^{-}_{2m,\alpha} -
S^{-}_{2m-1,\alpha}S^{+}_{2m,\alpha}\right)+
i\,d^{\alpha}_{+}\left(S^{+}_{2m,\alpha}S^{-}_{2m+1,\alpha} -
S^{-}_{2m,\alpha}S^{+}_{2m+1,\alpha}\right)\nonumber\\
&+& 2 S^{z}_{2m,\alpha}\left(S^{z}_{2m-1,\alpha}
+S^{z}_{2m+1,\alpha}\right)\Big] \, ,
\end{eqnarray}
where $S^{\pm}_{n,\alpha}=S^{x}_{n,\alpha}\pm S^{y}_{m,\alpha}$.
Thus,  at $d_{+}^{\alpha} \neq d_{-}^{\alpha}$, the translation symmetry of a $\alpha$-leg Hamiltonian is broken by the presence of nonequal amplitudes
of the spin current operator
$$
j^{\alpha}_{sp} \sim i \left(S^{+}_{n,\alpha}S^{-}_{n+1,\alpha} -
S^{-}_{n,\alpha}S^{+}_{n+1,\alpha}\right)\, .
$$
on odd and even links of a leg.

The next useful step is to rewrite the Hamiltonian
(\ref{Hamiltonian}) in a physically more suggestive manner by
rotating spins and gauging away the DM interaction term in each leg.
Here we follow the route, developed in the Ref. \cite{AJR_19}, in
the case of a single chain with alternating DM interaction. By
performing a site-dependent rotation around the $z$ axis of spins
along the leg $\alpha$  with relative angle $\vartheta^{\alpha}_{-}
$ for spins at consecutive odd-even sites ($2m-1,2m$) and
$\vartheta^{\alpha}_{+}$ for spins at consecutive even-odd sites
($2m,2m+1$), we introduce new spin variables
$\boldsymbol{\tau}_{2m,\alpha}$ and
$\boldsymbol{\tau}_{2m+1,\alpha}$ by
\begin{eqnarray}
\tau^+_{2m-1,\alpha} &=& e^{-i(m-1)(\vartheta^{\alpha}_{-}
+\vartheta^{\alpha}_{+} )}\,\,S^{+}_{2m-1,\alpha}\,
,\label{transfo_1}\\
\tau^{+}_{2m,\alpha} &=& e^{-im\vartheta^{\alpha}_{-}
-i(m-1)\vartheta^{\alpha}_{+} }\,\,
S^{+}_{2m,\alpha}\, ,\label{transfo_2}\\
\tau^z_{2m \pm 1,\alpha}&=& S^z_{2m \pm 1,\alpha}\,,\quad
\tau^z_{2m,\alpha}= S^z_{2m,\alpha}\ .\label{transfo_4}
\end{eqnarray}
Using (\ref{transfo_1})-(\ref{transfo_4})  we map the initial
$\alpha$-leg Hamiltonian onto
\begin{eqnarray}
\label{XXZ_w_Modulated_DMI-rotated} H_{leg}^{\alpha} &=& \frac{J}{2}
\sum_{m=1}^{N/2}\Big[\left(\cos\vartheta^{\alpha}_{-}
+d^{\alpha}_{-}\sin\vartheta^{\alpha}_{-} \right)
\left(\tau_{2m-1,\alpha}^{+}\tau_{2m,\alpha}^{-}+\tau_{2m,\alpha}^{-}\tau_{2m-1,\alpha}^{+}\right)
\nonumber\\
&&\hspace{10mm}+\left(\cos\vartheta^{\alpha}_{+}
+d^{\alpha}_{+}\sin\vartheta^{\alpha}_{+}
\right)\left(\tau_{2m,\alpha}^{+}\tau_{2m+1,\alpha}^{-}\, +\,
\tau_{2m,\alpha}^{-}\tau_{2m+1,\alpha}^{+}\right)\nonumber\\
&&\hspace{10mm}-i\left(\sin\vartheta^{\alpha}_{-}
-d^{\alpha}_{-}\cos\vartheta^{\alpha}_{-}
\right)\left(\tau_{2m-1,\alpha}^{+}\tau_{2m,\alpha}^{-}\, -\,
\tau_{2m,\alpha}^{-}\tau_{2m-1,\alpha}^{+}\right)\nonumber\\
&&\hspace{10mm} -i\left(\sin\vartheta^{\alpha}_{+}
-d^{\alpha}_{+}\cos\vartheta^{\alpha}_{+}
\right)\left(\tau_{2m,\alpha}^{+}\tau_{2m+1,\alpha}^{-}\, -\,
\tau_{2m,\alpha}^{-}\tau_{2m+1,\alpha}^{+}\right)  \,\nonumber\\
&&\hspace{10mm}+2
\tau_{2m}^{z}\,\left(\tau_{2m-1}^{z}+\tau_{2m+1}^{z} \right) \Big]\,
.
\end{eqnarray}
Choosing $\tan\vartheta^{\alpha}_{\pm} = d_{\alpha}^{\pm}$, gives
\begin{eqnarray}
\label{Cos-Sin_Theta} &&J(\sin\vartheta^{\alpha}_{\pm}   -
d_{\alpha}^{\pm}
\,\cos\theta_{\alpha}^{\pm}) = 0\,,\nonumber\\
 &&\hspace{-2mm} J(\cos\vartheta^{\alpha}_{\pm}   + d_{\pm}
\,\sin\theta_{\alpha}^{\pm}) = J
\sqrt{1+(d_{\alpha}^{\pm})^{2}}\,\equiv J_{\alpha}^{(\pm)}
\end{eqnarray}
and thus we eliminate the modulated DM interaction and obtain the
Hamiltonian of a each leg characterized by the alternating
transverse exchange interaction
\begin{eqnarray}
\label{Leg_w_Modulated_DMI_Rotated}H^{leg}_{\alpha}&=&
\sum_{m=1}^{N/2}\Big[\,\frac{J_{\alpha}^{(-)}}{2}\left(\tau^{+}_{2m-1,\alpha}\tau^{-}_{2m,\alpha}
+
\tau^{-}_{2m-1,\alpha}\tau^{+}_{2m,\alpha}\right)
+\,
\frac{J_{\alpha}^{(+)}}{2}\left(\tau^{+}_{2m,\alpha}\tau^{-}_{2m+1,\alpha}
+ \tau^{-}_{2m,\alpha}\tau^{+}_{2m+1,\alpha}\right)\nonumber\\
&&\hspace{10mm}+\, J \,\tau^{z}_{2m,\alpha}\left(
\tau^{z}_{2m-1,\alpha}+\tau^{z}_{2m+1,\alpha}\right)\Big]\, .
\end{eqnarray}
It is instructive to rewrite the Hamiltonian
(\ref{Leg_w_Modulated_DMI_Rotated}) in the following, more common,
form
\begin{eqnarray}
\label{Leg_w_Modulated_DMI_Rotated_v2} H^{leg}_{\alpha}&=&
J_{\alpha} \sum_{n}
\Big[\,\frac{1}{2}(1+(-1)^{n}\delta_{\alpha})\left(\tau^{+}_{n,\alpha}\tau^{-}_{n+1,\alpha}
+ \tau^{-}_{n,\alpha}\tau^{+}_{n+1,\alpha}\right) \,+
\gamma_{\alpha} \tau^{z}_{n,\alpha}\tau^{z}_{n+1,\alpha}\Big]\, ,
\end{eqnarray}
where, at $d^{\alpha}_{i}, \ll 1$ ( $i=\pm$),
\begin{eqnarray}
J_{\alpha} &=&\frac{J_{\alpha}^{(+)} + J_{\alpha}^{(-)}}{2}\simeq
J/\gamma_{\alpha}+{\cal O}\left((d^{\alpha}_{i})^{4}\right)\, ,
\\
 \delta_{\alpha}
&=&\frac{J_{\alpha}^{(+)}-J_{\alpha}^{(-)}}{J_{\alpha}^{(+)}+J_{\alpha}^{(-)}}\simeq
d^{\alpha}_{0}d^{\alpha}_{1}\gamma^{2}_{\alpha}+{\cal
O}\left((d^{\alpha}_{i})^{4}\right)\, \label{gamma-asterix}
\end{eqnarray}
and
\begin{eqnarray}\label{gamma-asterix}
&&\gamma_{\alpha} = J/J_{\alpha} =\frac{1}{
\sqrt{1+(d_{0}^{\alpha})^{2}+(d_{1}^{\alpha})^{2}}}\, .
\end{eqnarray}
Thus at $J_{\alpha}^{(+)} \neq J_{\alpha}^{(-)}$  the Hamiltonian
(\ref{Leg_w_Modulated_DMI_Rotated_v2}) is recognized as a
Hamiltonian of the $XXZ$ chain with easy-plane anisotrophy
($\gamma^{\ast}_{\alpha}<1$) and alternating transverse exchange.
Note that the alternation of the transverse exchange
$\delta_{\alpha} \neq 0$ only for finite $D^{\alpha}_{1}\neq 0$ and
$D^{\alpha}_{0} \neq 0$. In the following we will discard ${\cal
O}\left(d_{i}^{4}\right)$ corrections.

Gauging away of the DM interaction does not affect coupling with the
magnetic field and, as it can easily check by inspection, results
only via appearance of the site dependents phase factor in the
transverse part of the on-rung exchange. Inserting (\ref{transfo_1})
-(\ref{transfo_4}) in (\ref{InterLegCoup}) we obtain
\begin{eqnarray}
H_{rung} &=&  \sum_{n=1}^N \Big[\,\frac{J_{\bot}}{2} \left(
\,e^{-i\Phi_{n}}\,\tau^{+}_{n,1}\tau^{-}_{n,2}+h.c \right)+
J_{\bot}\tau^{z}_{n,1}\tau^{z}_{n,2} - H
\left(\tau^{z}_{n,1}+\tau^{z}_{n,2}\right)\Big] ,
\label{InterLegCoup_Rotated}
\end{eqnarray}
where
\begin{eqnarray}
&&\Phi_{2m-1} =
(m-1)(\vartheta_{1}^{+}+\vartheta_{1}^{-}-\vartheta_{2}^{+}-\vartheta_{2}^{-}),\\
&&\hspace{-3mm}\Phi_{2m} =
 m(\vartheta_{1}^{+}+\vartheta_{1}^{-}-\vartheta_{2}^{+}-\vartheta_{2}^{-})-
 (\vartheta_{1}^{+}-\vartheta_{2}^{+})\, .
\label{Phi_n}
\end{eqnarray}
In the particular case of a ladder with identical legs and in-phase
modulation of the DM interaction ($D_{1}(n)=D_{2}(n))$, \,
$\Phi_{2m-1}=0$ and $\Phi_{2m}=\vartheta^{1}_{+}-\vartheta^{2}_{+}$.

Thus, after gauging away the DM interaction the Hamiltonian under
consideration takes the following form
\begin{eqnarray}
\label{Hamiltinian_MDMI_Rotated} {\cal H}&=&
 \sum_{n,\alpha}
\Big[\,\frac{J_{\alpha}}{2}(1+(-1)^{n}\delta_{\alpha})
\left(\tau^{+}_{n,\alpha}\tau^{-}_{n+1,\alpha} +
\gamma^{\ast}_{\alpha}
\tau^{-}_{n,\alpha}\tau^{+}_{n+1,\alpha}\right)
 + \,\tau^{z}_{n,\alpha}\tau^{z}_{n+1,\alpha} - H
(\tau^{z}_{n,1}+\tau^{z}_{n,1})\Big]\nonumber\\
&+&J_{\bot}\sum_{n=1}^N \Big[\,\frac{1}{2} \left(
\,e^{-i\Phi_{n}}\,\tau^{+}_{n,1}\tau^{-}_{n,2}+h.c \right) +
\tau^{z}_{n,1}\tau^{z}_{n,2} \, \Big] .
\end{eqnarray}


\section{Effective Hamiltonian in the case of strong rung exchange and magnetic field}\label{sec3}

In what follows we restrict ourselves by consideration of the limit
of strong rung exchange and magnetic field ($J_{\bot},H \gg
J_{\alpha}$) and follow the route developed already decades ago to
study magnetic phase diagram of a two-leg ladder in this limit
\cite{Mila_98,Totsuka_98}.

We start from the case $J_{\alpha} = 0$, where the system decouples
into a set of noninteracting rungs in a magnetic field
\begin{eqnarray}
\label{H-Rungs}{\cal H} &= & \sum_{n=1}^N \Big[\,\frac{J_{\bot}}{2}
\left( \,e^{-i\Phi_{n}}\,\tau^{+}_{n,1}\tau^{-}_{n,2}+h.c \right) +
J_{\bot}\tau^{z}_{n,1}\tau^{z}_{n,2} - H
(\tau^{z}_{n,1}+\tau^{z}_{n,2})\, \Big]
\end{eqnarray}
and all eigenstates of the Hamiltonian (\ref{H-Rungs}) can be
written as a product of rung states. The phase factor $\Phi_{n}$ does not change the
eigenvalues of a pair of coupled spins on a rung -- at each rung,
two spins form either a singlet state $|s\ra$ with energy $E_{s} =
-0.75J_{\bot}$ or in one of the triplet states $|t^{+}\ra$,\,
$|t^{0}\ra$ and $|t^{-}\ra$ with energies $E_{t^{+}} = 0.25J_{\bot}
- H$, $E_{t^{0}} = 0.25J_{\bot}$ and $E_{t^{-}}(n)=0.25J_{\bot} +
H$, respectively. When $H$ is small, the ground state consists of a
product of rung singlets. As the field $H$ increases, the energy of
the triplet state $|t^{+}\rangle$ decreases and at $H \simeq
J_{\perp}$ forms, together with the singlet state, a doublet of
almost degenerate low energy state, split from the remaining high
energy two triplet states. To project the Hamiltonian (\ref{Hamiltinian_MDMI_Rotated})
on the corresponding  low-energy sector we introduce the effective
spin operator $\sigma$ which act on these states as \cite{Mila_98}
\begin{eqnarray}
&&\sigma_{n}^{z}|\,s_{0}>_{n}~ = -\frac{1}{2}|\,s_{0}>_{n}\, , ~~~~
\sigma_{n}^{z}|\,t^{+}>_{n} ~ = \frac{1}{2}|t^{+}>_{n}\, ,\nonumber \\
&&\sigma_{n}^{+}|\,s_{0}>_{n} ~ = ~~~|\,t^{+}>_{n}\, , ~~~~~
\sigma_{n}^{+}|t^{+}>_{n}~ = ~~~0 \, , \\
&&\sigma_{n}^{-}|\,s_{0}>_{n} ~ = ~~~~ 0 \, ,~~~~~~~~~~~~
\sigma_{n}^{-}|\,t^{+}>_{n} = |\,s_{0}>_{n}\, . \nonumber
\end{eqnarray}
The relation between the initial ladder spin operator
${\mbox{\boldmath $\tau$}}_{n}$  and the pseudo-spin operator
${\mbox{\boldmath $\sigma$}}_{n}$ in this restricted subspace can be
easily derived by inspection,
\be \tau^{\pm}_{n,\alpha} =
(-1)^{\alpha}\frac{1}{\sqrt{2}}\sigma^{\pm}_{n}\, , \quad
\tau^{z}_{n,\alpha} =
\frac{1}{2}\left(\frac{1}{2}+\sigma^{z}_{n}\right) \, .
\label{S-Tau-relations} \ee
Using (\ref{S-Tau-relations}), to the first order and up to a
constant, we easily obtain the following effective Hamiltonian
\begin{eqnarray}
\label{eqn:7} \mathcal{H}_{eff} &=& J^{\ast}\sum_{n=1}^{N} \big[
\frac{1}{2}(1 + (-1)^n \delta^{\ast})
(\sigma_{n}^{+}\sigma_{n+1}^{-} + \sigma_{n}^{-}\sigma_{n+1}^{+} )+
\gamma^{\ast}\,\sigma_{n}^{z}\sigma_{n+1}^{z}  - h\,
\sigma_{n}^{z}\big]
\end{eqnarray}
which describes a single spin 1/2 chain with a fixed XY anisotropy
of $\gamma^{\ast}$ and alternating transverse exchange in an
effective uniform magnetic field $H_{eff}$. Here
\begin{eqnarray}\label{eqn:8a}
J^{\ast} & = &\frac{J_{1}+J_{2}}{2}\, , \\
\delta^{\ast} & =&
\frac{\delta_{1}J_{1}+\delta_{2}J_{2}}{J_{1}+J_{2}}\, ,\label{eqn:8b} \\
\gamma^{\ast} &
=&\frac{1}{2}\frac{\gamma_{1}J_{1}+\gamma_{2}J_{2}}{J_{1}+J_{2}}\label{eqn:8c}
\end{eqnarray}
and
\begin{eqnarray}\label{eqn:9}
H_{eff} = J^{\ast}\,h & = & H - J_{\perp} - \gamma^{\ast}J^{\ast}\,.
\end{eqnarray}
It is worth to notice that a similar problem has been studied
intensively in past years
\cite{Chitra_Giamarchi_97,Cabra_Honecker_97-98,Derzhko_07,Dmitriev_Krivnov_12}.

In absence of the DM interaction, ($\delta_{\alpha}=0$ and
$\gamma_{\alpha}=1$) the Hamiltonian (\ref{eqn:7}) is the
Hamiltonian of XXZ chain with a fixed XY anisotropy of $1/2$ and
coincides with the corresponding effective Hamiltonian for a
standard ladder derived in the strong-rung and limit \cite{Mila_98}.
As it follows from (\ref{eqn:8b}) in the strong on-rung coupling
limit the effect of modulated DM interaction, displayed via the
alternation of the transverse exchange, is most pronounced in the
case of in-phase modulation of the DM coupling in both legs, where
$\delta_{1}$ and $\delta_{2}$ are of the same sign and is the
weakest in the case of anti-phase modulation where $\delta_{1}$ and
$\delta_{2}$ are of different signs.


\section{Magnetic properties}

\subsection{The first critical field $H_{on}$ and the saturation field $H_{sat}$}

The performed mapping allows to determine critical fields $H_{on}$
corresponding to the onset of magnetization in the system and the
saturation field $H_{sat}$ \cite{Mila_98}. The easiest way to
express $H_{on}$ and $H_{sat}$ in terms of ladder parameters is to
perform the Jordan-Wigner transformation which maps the problem onto
a system of interacting spinless fermions \cite{JW_1928}:
\begin{eqnarray}\label{eqn:9}
&& \sigma_{n}^{-} = e^{i\pi\sum_{m<n} c_{n}^{\dag}c_{n}}\, c_{n},\nonumber\\
&& \sigma_{n}^{+} =  c_{n}^{\dag}\, e^{-i\pi\sum_{m<n} c_{n}^{\dag} c_{n}}, \\
&& \sigma_{n}^{z} = c_{n}^{\dag} c_{n} - 1/2 \equiv
\rho_{n},\nonumber
\end{eqnarray}
the model (\ref{eqn:7}) could be rewritten in the following form:
\begin{eqnarray}
H_{sf} & = & t\sum_{n}(1 + (-1)^n \delta^{\ast})(a_{n}^{+}a_{n+1} +
h.c.) + V \sum_{n}\rho_{n}\rho_{n+1} - \mu\,\sum_{n} \rho_{n},
\label{Hamiltonian_SpFrm}
\end{eqnarray}
where $t=V=J^{\ast}/2$ and $\mu = H - J_{\perp}$. The lowest
critical field $H_{on}$ corresponds to that value of the chemical
potential $\mu_{c}$ for which the band of spinless fermions starts
to fill up. In this limit we can neglect the interaction term in Eq.
(\ref{Hamiltonian_SpFrm}) and obtain the model of free massive
particles with spectrum
$$
E^{\pm}(k)=-\mu \pm J^{\ast}(cos^{2}(k) +
\delta^{\ast\,2}\sin^{2}(k))^{1/2}.
$$
This gives
\begin{eqnarray}
H_{on} = J_{\perp} -J^{\ast} \, .
\end{eqnarray}
Thus at $H<H_{on}$ the effective spin-chain model is fully polarized
with magnetization (per site) $m=-1/2$. This corresponds to the
state, where on all rungs pairs of spins form singlets and
respectively, as it follows from (\ref{S-Tau-relations}) the net
magnetization (per site) of the of the initial ladder system
$$
M=\frac{1}{N}\sum_{n,\alpha}\tau^{z}_{n,\alpha}=0\, .
$$

A similar argument can be used to determine $H_{sat}$. In the limit
of almost saturated magnetization, it is useful to make a
particle-hole transformation and estimate $H_{sat}$ from the
condition where the transformed hole band starts to fill, what gives
\begin{eqnarray}
H_{sat} = J_{\perp} +J^{\ast}.
\end{eqnarray}
Therefore for $H>H_{sat}$ the effective spin-chain model is in the
fully polarized state with $m=1/2$, all on-rung pairs of spins form
triplets and respectively, as it follows from
(\ref{S-Tau-relations}) the net magnetization of the of the initial
ladder system
$$
M=\frac{1}{N}\sum_{n,\alpha}\tau^{z}_{n,\alpha}= 1 \, .
$$

\subsection{Magnetization plateau at  $M=0.5$}

To characterize excitation spectrum and magnetic properties of the
system at intermediate values of the magnetic fields $H_{on}< H <
H_{sat}$ we use the continuum-limit bosonization treatment of the
model (\ref{eqn:7}). Following the usual procedure in the low energy
limit, we bosonize the spin degrees of freedom at fixed
magnetization $m$ and the interaction term becomes ~\cite{GNT}
\bea \sigma_{n}^{z}  &=&  m + \sqrt{\frac{K}{\pi}} \partial_x
\phi(x) +  \frac{A_1}{\pi} \sin\left( \sqrt{4\pi K}\phi(x)
+(2m+1)\pi n \right) \, ,
\label{bosforSigma_z}\\
\sigma_{n}^{+}&=&e^{-i\sqrt{\pi/K}\theta(x)}+
\frac{B_1}{\pi}e^{-i\sqrt{\frac{\pi}{K}}\theta(x)}\sin\left(\sqrt{4\pi
K}\phi(x)+(2m+1)\pi n\right), \label{bosforSigma_+} \eea
where $A_1$ and $B_1$ are  non-universal real constants of the order
of unity \cite{Hikihara01}. and $m$ is the magnetization (per site)
of the chain. Here $\phi(x)$ and $\theta(x)$ are dual bosonic
fields, $\partial_t \phi = v_{s}
\partial_x \theta $, and satisfy the following commutation
relations
\begin{eqnarray}
\label{regcom}
&& [\phi(x),\theta(y)]  = i\Theta (y-x)\,,  \nonumber\\
&& [\phi(x),\theta(x)]  =i/2\, ,
\end{eqnarray}
and $K(\gamma^{\ast},m)$ is the spin-stiffness parameter for a chain
with anisotropy $\gamma^{\ast}$ and magnetization $m$. At zero
magnetization \cite{LP_75}
\begin{equation}
K(\gamma^{\ast},0)=\frac{\pi}{2\left(\pi-\arccos\gamma^{\ast}
\right)} \, .\label{K}
\end{equation}

Using (\ref{bosforSigma_z}) and (\ref{bosforSigma_+}) we get the
following bosonized Hamiltonian
\begin{eqnarray}  \label{H_XXZ+D1_bos}
{\cal H} &= & u\int dx
\,\Big[\frac{1}{2}(\partial_{x}\phi)^2+\frac{1}{2}(\partial_{x}\theta)^2
- h\,\sqrt{\frac{K}{\pi}}\partial_x\phi
+\frac{\delta^{\ast}}{2\pi^{2}\alpha^{2}} \cos\left(\sqrt{4\pi
K}\phi + 2\pi m x\right)\, \Big]\, ,
\end{eqnarray}
where $u \simeq J^{\ast}/K$. In deriving (\ref{H_XXZ+D1_bos}) the
strongly irrelevant at $\gamma^{\ast}<1/2$ (i.e for $K>3/4$) term
$\sim \cos\left(\sqrt{16\pi K}\phi+2\pi m x\right)$ has been
omitted.

As we observe, for arbitrary $m \neq 0$ the cosine term in
(\ref{H_XXZ+D1_bos}) contains oscillating factor and therefore has
to be ignored in the continuum limit. Therefore in this case the
effective spin chain model is critical and its long-wavelength
excitations are described by the standard Gaussian theory  and the
magnetic field term can be easily absorbed by a shift
$\partial_{x}\phi \rightarrow \partial_{x}\phi) +\frac{K}{\pi}h $
resulting to linear in $h$ magnetization of a gapless critical phase
\cite{shift}.

At $m=0$ i.e. $M=0.5$ the oscillating factor in the cosine term is
absent and the continuum-limit bosonized version of the effective
spin-chain model is given by the Hamiltonian
\begin{eqnarray}  \label{H_XXZ+D1_bos_CIC_T}
{\cal H}&=&u\int dx
\,\Big[\frac{1}{2}(\partial_{x}\phi)^2+\frac{1}{2}(\partial_{x}\theta)^2
- h\,\sqrt\frac{K}{\pi}\partial_x\phi
 +\frac{\delta^{\ast}}{2\pi^{2}\alpha^{2}}
\cos\sqrt{4\pi K}\phi \, \Big]\, .
\end{eqnarray}
The Hamiltonian (\ref{H_XXZ+D1_bos_CIC_T}) is easily recognized as
the standard Hamiltonian for the commensurate-incommensurate phase
transition \cite{C_IC_transition_1,C_IC_transition_2} which has been
intensively studied in the past using bosonization approach
\cite{JN_79} and the Bethe ansatz technique \cite{JNW_84}. Below we
use the results obtained in \cite{JN_79,JNW_84} to describe
magnetization plateau and the transitions from a gapped (plateau) to
gapless paramagnetic phases.

Let us first consider $h=0$. In this case, the continuum theory of
the initial spin-chain model in the magnetic field $H=J_{\perp} +
\gamma^{\ast}J^{\ast}$ is given by the quantum sine-Gordon (SG)
model with a massive term $\sim \delta^{\ast}\cos(\sqrt{4\pi
K}\phi)$ with $3/4\leq K <1$. From the exact solution of the SG
model~\cite{DHN_74,KF_75,Zamolodchikov_95}, it is known that at
$3/4\leq K <1$ the excitation spectrum of the model  consists of
solitons and antisolitons with masses \cite{Zamolodchikov_95}
\be \Delta = J^{\ast}{\cal C}(K)
\left(\delta^{\ast}\right)^{1/(2-K)} \, ,
\label{SG-mass_Zamolodchikov} \ee
where ${\cal C}(K)$ is a constant of the order for all $K$ from the
considered sector  $0.75 \leq K <1$.
$$
{\cal C}(K)\!\!=\!\! \frac{2}{\sqrt{\pi}}
\frac{\Gamma(\frac{K}{8-2K})}{\Gamma(\frac{2}{4-K})}
\left[\frac{\Gamma(1-K/4)}{2\Gamma(K/4)}\right]^{2/(4-K)}\!\!.
$$
and $\Gamma$ is the Gamma function.

In the gapped phase, the ground state properties of the system are
determined by the dominant potential energy term $\sim
\cos(\sqrt{4\pi K}\phi)$ and therefore in the gapped phases, the
field $\phi$ is pinned in one of the vacua
\be \la 0| \sqrt{4\pi K}\phi|0 \ra = (2n+1)\pi \, , \label{minima}
\ee
to ensure the minimum of the energy. From
(\ref{bosforSigma_z})-(\ref{bosforSigma_+}) it is clearly seen in
the ordered phase local spin degrees of freedom are fully suppressed
and only the on link dimer order
$$
D_{n}=(-1)^{n}\left(\sigma_{n}^{+}\sigma_{n+1}^{-} +
\sigma_{n}^{-}\sigma_{n+1}^{+}\right)\sim \cos(\sqrt{4\pi K}\phi)
$$
show a long range order. In terms of initial ladder system it
corresponds to the formation by singlet and triplet located on
neighboring rungs of a plaquette of the coherent entangled pair.

At $h \neq 0$ (i.e. $H \neq J_{\perp} + \gamma^{\ast}J^{\ast}$), the
presence of the gradient term in the Hamiltonian
(\ref{H_XXZ+D1_bos_CIC_T}) makes it necessary to consider the ground
state of the sine-Gordon model in sectors with nonzero topological
charge. The effective chemical potential
$$\sim h_{eff}^{0}\sqrt{\frac{K}{\pi}}\partial_{x}\phi$$
 tends to
change the number of particles in the ground state i.e. to create
finite and uniform density of solitons. It is clear that the
gradient term in (\ref{H_XXZ+D1_bos_CIC_T}) can be again eliminated
by a gauge transformation
$$\phi \rightarrow \phi +
h_{eff}^{0}\sqrt{\frac{K}{\pi}}\,x \, ,$$
however this immediately implies that the vacuum distribution of the
field $\phi$ will be shifted with respect of the minima
(\ref{minima}). This competition between contributions of the smooth
and modulated components of the magnetic field is resolved as a
continuous phase transition from a gapped state at $|h| < \Delta$ to
a gapless (paramagnetic) phase at $|h|
> \Delta$, where $\Delta$ is the soliton mass~\cite{C_IC_transition_1}. This
condition gives two critical values of the magnetic field for each
plateau
\bea H_{c}^{\pm} &=&  J_{\perp} + \gamma^{\ast}J^{\ast} \pm
J^{\ast}{\cal C}(K) \left(\delta^{\ast}\right)^{1/(2-K)}\eea
and respectively determines the width of the magnetization plateau
by
\be H_{c}^{+}- H_{c}^{-} \simeq
2J^{\ast}\left(\delta^{\ast}\right)^{1/(2-K)} \, . \label{Platou}
\ee

As usual in the case of C-IC transition, the magnetic susceptibility
of the system shows a square-root divergence at the transition
points:
$$
\chi(H) \sim \sqrt{(H/H_{c}^{-})-1} \quad {\mbox for}\quad H \geq
H_{c}^{-}$$
and
$$\chi(H)=\sqrt{1-(H/H_{c}^{+})}  \quad {\mbox for}\quad   H \leq H_{c}^{+} \, .
$$

Thus we obtain the following magnetic phase diagram for a ladder
with alternating DMI along the legs. For $H \leq H_{on}$, the system
is in a rung-singlet phase with zero magnetization and vanishing
magnetic susceptibility. For $H > H_{on} $   some of singlet rungs
melt and the magnetization increase as $ M \sim \sqrt{(H/H_{on})
-1}$ . With further increase of the magnetic field the system
gradually crosses to a regime with linearly increasing
magnetization. However, in the vicinity of the magnetization plateau
at $M=0.5$, for $H \leq H_{c}^{-}$   this linear dependence changes
again into  a square-root behavior $M=0.5 - \sqrt{1-(H/H_{c}^{-})}$.
For fields in the interval between $H_{c}^{-}<H<H_{c}^{+}$ the
magnetization is constant $M=0.5$. At $H>H_{c}^{+}$the magnetization
increases as $M=0.5 + \sqrt{(H/H_{c}^{+})-1}$, then passes again
through a linear regime until, in the vicinity of the saturation
field $H<H_{sat}$ , it becomes $M=1 - \sqrt{1-(H/H_{sat})}$.

\section{Summary} We have studied the ground state magnetic
phase diagram of a spin $S=1/2$ antiferromagnetic two-leg ladder in
the case where spins along the legs are affected by the modulated
with period of two lattice units Dzyaloshinskii-Moriya interaction.
The model is studied in the limit of strong rung exchange and
magnetic field. It is shown, that the very presence of modulated DM
term opens a gap in the excitation spectrum of the ladder at
magnetization equal to half of its saturated value $M_{sat}$. The
value of a gap is determined simultaneously by the uniform and
staggered components of the DMI and is absent if one of these
components is zero.  Respectively, at $M=0.5M_{sat}$ the plateau at
magnetization curve appears of a width which determined by the gap
in the excitation spectrum. We have also shown, that in the gapped
phase unconventional magnetic order, where triplet and singlet
states localized on neighboring rungs form an entangled dimer states
and this dimerized plaquettes form a long-range order phase with
period two lattice sites wavelength.

To conclude we have to stress that the obtained phase diagram is not
a particularity of the considered case of two lattice unit
modulation of the DMI. The gapped phases, characterized by
magnetization plateau and  dimerized plaquette order, appear in the
case of arbitrary, but commensurate with the lattice unit modulation
of the DMI interaction $D_{\alpha}(n)=D_{0}^{\alpha}+
\cos(qn)D_{1}^{\alpha}$ at corresponding values of the magnetization
$M=p/q M_{sat}$, where $p$ and $q$ are natural numbers and $p<q$.

\section{Acknowledgment}

We are grateful to Prof. A.A. Nersesyan for his interest
in this work and helpful comments.

\vspace{0.3cm}

\end{document}